\documentclass[10pt,aps,prr,final,twocolumn,amsmath,amssymb,superscriptaddress,longbibliography]{revtex4-1}
\usepackage[pdftex, colorlinks=true, linkcolor=blue, citecolor=blue, urlcolor=blue]{hyperref}
\usepackage{graphicx}
\usepackage{aurical}
\usepackage[T1]{fontenc}
\usepackage[version=3]{mhchem}

\graphicspath{{./figures/}}
\DeclareGraphicsExtensions{.png,.pdf}

\newcommand{\una}{Universit\'{e} de Nouakchott, Facult\'{e} des Sciences et Techniques, D\'{e}partement de Physique, Avenue du Roi Fai\c{c}al, 2373, Nouakchott, Mauritania}
	
\begin{document}
\title{Comment on "Charge pumping with strong spin-orbit coupling: Fermi surface breathing, Berry curvature, and higher harmonic generation"}
\author{O. Ly}
\email{ousmane.ly.physics@gmail.com}
\affiliation{\una}
	
\maketitle
We have recently proposed the emergence of high-order harmonics from magnetic dynamics in the presence of spin-orbit coupling (SOC) \cite{Ly2022}. While the initial proposal has been carried out in a non-perturbative framework \cite{tkwant} with arbitrary exchange coupling $\Delta$, a subsequent study has confirmed the effect in the low coupling regime \cite{Manjarres2021}, where only few harmonics could be observed. The scaling up of this effect in terms of $\Delta$, the precession angle $\theta$ and the SOC strength has been so far studied \cite{Ly2023c} as well as the counterpart of the effect in the presence of topological orders instead of SOC \cite{Ly2023}.

Furthermore, the highly nonlinear response underlying the effect has been recently attributed to the highly nonlinear dynamics of the instantaneous energy levels \cite{Ly2024}.

In their recent work \cite{Manchon2024}, Manchon and Pezo develop a Keldysh Green's function based framework for the study of this effect. In this study the authors assume that their response formula is obtained up to a linear order in the magnetic dynamics $\mathbf{m}(t)=(\cos \theta, \sin \theta \sin \omega t, -\sin \theta \cos \omega t)$ (as stated above Eq.(3)). This means that their theory would be just linear in $\sin \omega t$ and/or $\cos \omega t$. Therefore, it would in any sense be capable of displaying higher-order harmonics.
Consequently, their assumed adiabatic theory cannot be used to study high harmonic generation.

A correct derivation of the adiabatic current should rather be highly nonlinear in the driving dynamics $\mathbf{m}(t)$ in order to enable for high harmonic generation.

To illustrate this, one starts from the expression of the Berry curvature obtained by the authors themselves in Eq(23). Where the standard adiabatic theory \cite{Xiao2010} is used.
The Berry curvature is found to depend on $\lambda_k=\sqrt{\Delta^2+\alpha_R^2 k^2+2\Delta\alpha_R(\mathbf{p}\times \mathbf{z}).\mathbf{m}(t) }$. And thereby depends highly non-linearly on $\mathbf{m}(t)$, leading to the emergence of high harmonics in the underlying adiabatic current. 

We note that $\lambda_k$ corresponds to the dynamical term $\sqrt{D}$ that was recently introduced in ref. \cite{Ly2024} as the main precursor for high harmonic generation in Rashba systems subjected to magnetic dynamics. 

Finally, we highlight that the fact that the theory developed in ref. \cite{Manchon2024} is linear in $\mathbf{m}(t)$ implies that it is also linear in the exchange coupling $\Delta$. The later appears similarly in a highly nonlinear fashion within the expression of the adiabatic Berry curvature discussed above. In contrast, it will only appear linearly in the transport equation of \cite{Manchon2024} (Eq.(7)).

This demonstrates clearly that the theory (linear in $\mathbf{m}(t)$) presented in \cite{Manchon2024} cannot be used to describe the physics of high harmonic generation, which, by definition, is of nonlinear nature. 

\bibliography{refs}
\end{document}